\pdfoutput=1 
\documentclass{JINST} 
\usepackage{graphicx}

\title{Combined readout of a triple-GEM detector}

\author{
V. C.~Antochi$^{a}$,
E.~Baracchini$^{a}$,
G.~Cavoto$^{a,b}$,
E.~Di Marco$^{a}$,
M.~Marafini$^{a,c}$, 
G.~Mazzitelli$^{d}$,
D.~Pinci$^{a}$\thanks{Corresponding author.},  
F.~Renga$^{a}$,
S.~Tomassini$^{d}$ and
C.~Voena$^{a}$\\
\llap{$^a$}Istituto Nazionale di Fisica Nucleare\\  
Sezione di Roma, I-00185, Italy\\ 
\llap{$^b$}Dipartimento di Fisica\\  
Sapienza Universit\`a di Roma, I-00185, Italy\\ 
\llap{$^c$}Museo Storico della Fisica e Centro Studi e Ricerche "Enrico Fermi" \\ 
Piazza del Viminale 1, Roma, I-00184, Italy\\ 
\llap{$^d$}Istituto Nazionale di Fisica Nucleare \\  
Laboratori Nazionali di Frascati, I-00040, Italy\\ 
 
E-mail: \email{davide.pinci@roma1.infn.it}}

\abstract{
Optical readout of GEM based devices by means of high granularity and
low noise CMOS sensors allows to obtain very interesting tracking performance.
Space resolution of the order of tens of $\mu$m were measured on the GEM plane
along with an energy resolution of 20\%$\div$30\%.
The main limitation of CMOS sensors is represented by their poor information about
time structure of the event.
In this paper, the use of a concurrent light readout by means of a suitable 
photomultiplier and the acquisition of the electric signal induced on the
GEM electrode are exploited to provide the necessary timing informations. 
The analysis of the PMT waveform allows a
3D reconstruction of each single clusters with a 
resolution on {\it z} of 100 $\mu$m.
Moreover, from the PMT signals it is possible to obtain 
a fast reconstruction of the energy released within 
the detector with a resolution
of the order of 25\% even in the tens of keV range
useful, for example, for triggering purpose.}

\begin{document}

\section*{Introduction}

Time Projection Chambers based on a Micro Pattern Gaseous Detector
optical readout represent ideal candidates for high resolution
particle tracking.
MPGD are a simple solution for equipping large surfaces, 
ensuring very good spatial and timing resolution.
In particular Gas Electron Multipliers \cite{bib:gem} 
are very easy to assemble and their structure 
is able to suppress the Ion Back Flow 
inside the sensitive volume. 
In these devices, during the multiplication processes, 
photons are produced along with electrons 
by the gas through atomic and molecular de-excitation.
Optical readout of gas detectors offers several advantages:

\begin{itemize}
\item optical sensors are able to provide high granularities 
along with very low noise levels and high sensitivities;
\item optical coupling allows to keep sensors out of the sensitive 
volume (no interference with HV operation and lower gas contamination);
\item suitable lens allow to acquire large surfaces with small sensors.
\end{itemize}

This approach was already studied in the past 
(\cite{bib:fraga1}, \cite{bib:fraga2}).
In last years, the developments of large granularity and low noise
CMOS sensors made it possible to obtain very interesting results with
detectors based on optically readout GEM (\cite{bib:orange1}, \cite{bib:orange2}).
Very good tracking performance were obtained
with a space resolution (on the GEM plane) of 35 $\mu$m even for minimum ionising particles.
Moreover, they allow the measurement not only of the total released energy but also of the energy 
release density along the tracks that can be very useful 
for particle identification (\cite{bib:orange3}).
To overcome the poor timing information provided by the CMOS,
a combined light readout with a fast PMT is proposed.
In this paper, the performance obtained by reading out the light
provided by a triple-GEM structure with a CMOS sensor together with
a fast PMT, are described. 

\section{Experimental set-up}

\begin{figure}
\centering
\includegraphics[width=.95\textwidth]{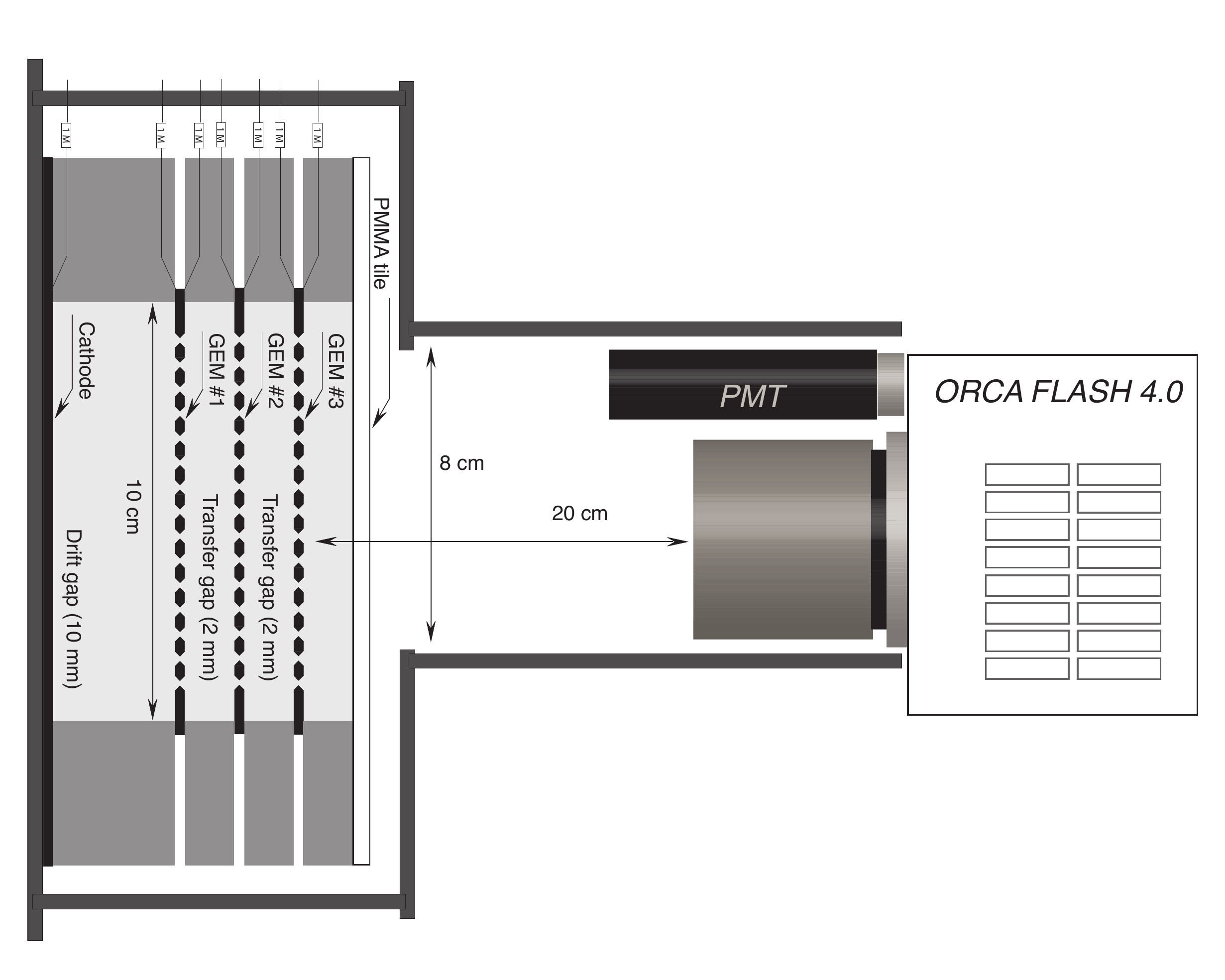}
\caption{Drawing (not to scale) of the experimental setup.
In particular, the CMOS camera, with its lens, 
and the PMT are visible.}
\label{fig:setup}
\end{figure}

A sketch of the setup used to test 
ORANGE (Optically ReAdout GEm) prototype 
(described in more details in \cite{bib:orange3})
is shown in Fig. \ref{fig:setup}.

Measurements were taken with 450 MeV electrons  
at the ``Beam Test Facility'' \cite{bib:btf}
of INFN Laboratori Nazionali di Frascati.

Three standard $10\times10$ cm$^2$ GEM were used 
with 2 millimetre transfer gaps between the GEM and a 
drift gap of 1 cm. 
The sensitive volume was, therefore, 0.1 litre. 
The device was flushed with an He/CF$_4$ 60/40 mixture
and the gas volume was closed by means of a transparent
PMMA window.
The detector was operated with a drift field of 1.5 kV/cm
and two transfer fields of 2.0 kV/cm.
In this configuration an electron velocity 
in the drift gap of of about 74 $\mu$m/ns was
estimated with Garfield \cite{bib:garfield}.

\subsection{Detector readout}

\begin{figure}
\centering
\includegraphics[width=.95\textwidth]{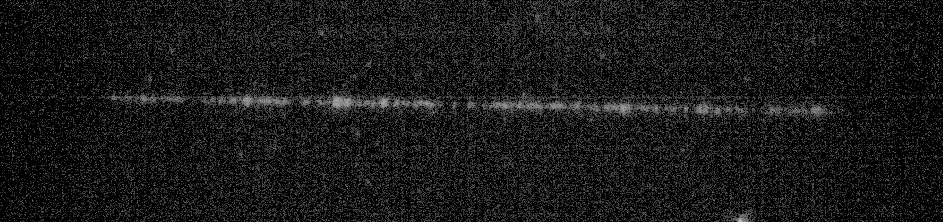}
\caption{Examples of acquired image of a 450 MeV electron.}
\label{fig:e_btf}
\end{figure}

The light emitted by gas mixture in the third GEM channels was 
readout by an Orca Flash 4 CMOS-based 
camera\footnote{For more details visit the site www.hamamatsu.com}
equipped with a large aperture (f/0.95) lens.
Very clear images of high energy electrons from the 
beam as the one shown in 
Fig. \ref{fig:e_btf}, were acquired by this device.

\begin{figure}
\centering
\includegraphics[width=.95\textwidth]{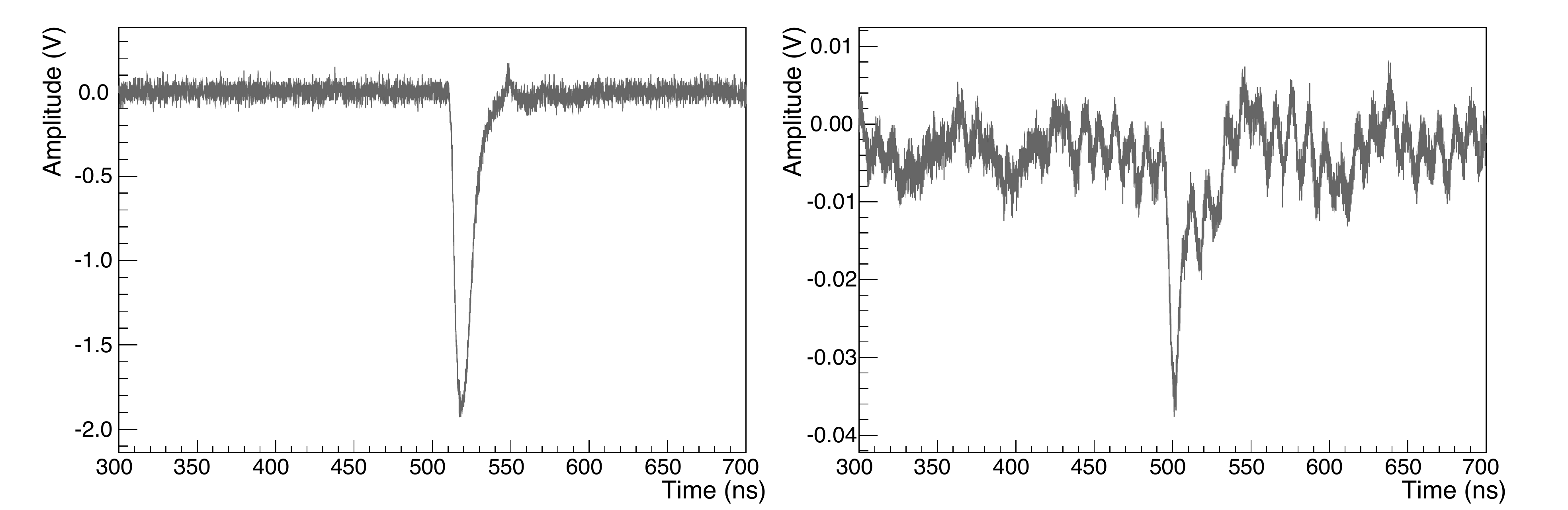}
\caption{Example of an acquired waveform of the PMT output and the 
electric signal induced on the bottom electrode of the third GEM (G3D).}
\label{fig:pmt-g3d}
\end{figure}

The image in Fig. \ref{fig:e_btf} represents
a single electron traveling
in the drift gap, parallel to the GEM foils.
In order to acquire the time structure of the signals, 
light was concurrently readout by a 25-mm diameter 
PMT\footnote{Hamamatsu H10580} placed close
to the camera lens.
Moreover, the electric signal
induced by the motion of the electrons on the third GEM bottom electrode (G3D) was
acquired by means of a 10 GS/s sampling oscilloscope\footnote{Lecroy Waverunner 7300}.
These two signals are able to provide two independent 
measurements of time structure of the event.
The acquired waveform
of the PMT output
and of the electrical signal 
in the event of in Fig. \ref{fig:e_btf}
are shown respectively on the left
and on the right of Fig. \ref{fig:pmt-g3d} 

\begin{figure}
\centering
\includegraphics[width=.95\textwidth]{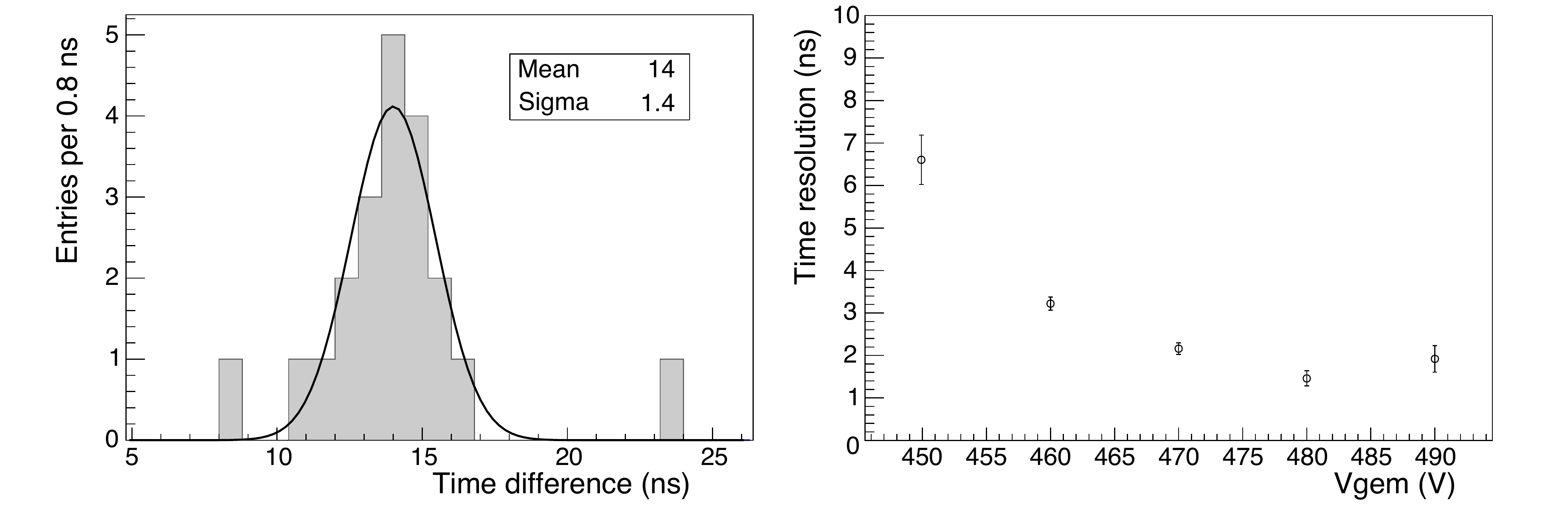}
\caption{Distribution of the differences of the arrival time of the
PMT and G3D signals with a superimposed gaussian fit (left) and of the 
behaviour of the time resolution as a function of the GEM voltage supply (right).}
\label{fig:trex}
\end{figure}

In both cases clear and narrow (less than 10 ns FWHM) 
waveforms are visibile. The relative noise level in the electric
signal is considerably larger, most likely due to jitter on the high 
voltage supply line.
From a gaussian fit of the PMT output,
a sigma of 5.5 ns was measured.
By measuring the distribution of the differences of the time of arrival of the two signals,
the time resolution of the detector for tracks parallel to the GEM plane
could be evaluated (Fig. \ref{fig:trex}, left).

The behaviour of the time difference as a function of the voltage applied to the
three GEM foils (Fig. \ref{fig:trex}, right) shows that a sigma of 
less than 2 ns can be achieved.
Assuming that the time jitter of the signal is almost the same, 
a time resolution of 2/$\sqrt{2}$ ns can be evaluated for both of them.
By using the calculated drift velocity,
it is thus possible to conclude that the absolute track {\it z}
position can be reconstructed with a resolution of about 100 $\mu$m.

\subsection{Exploiting the time information: combined readout}

\begin{figure}
\centering
\includegraphics[width=.95\textwidth]{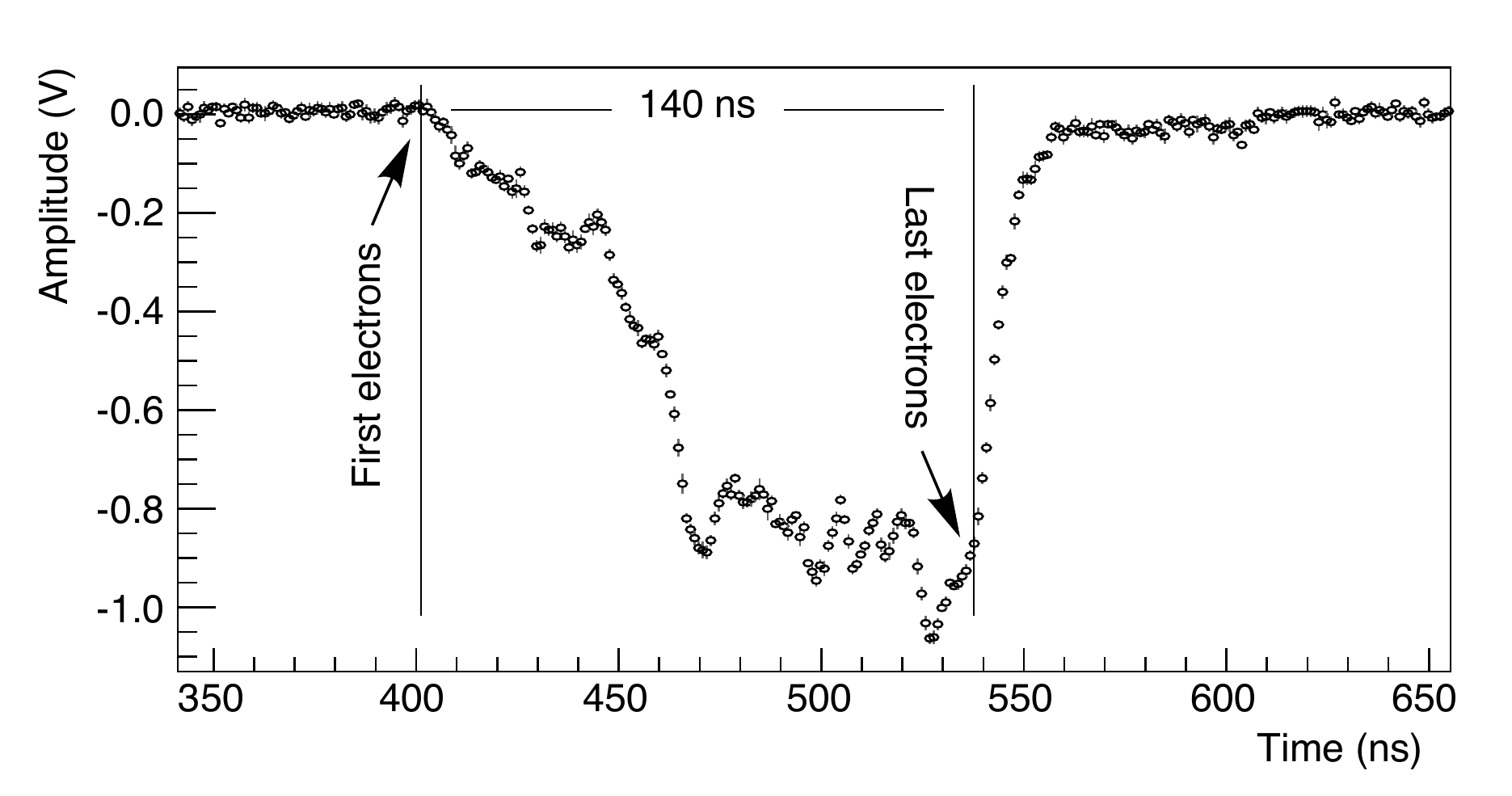}
\caption{PMT waveform 
for a track crossing the drift gap inclined with respect to the GEM plane.}
\label{fig:pmt-inclined}
\end{figure}

In Fig. \ref{fig:pmt-inclined}, the PMT signal 
is shown for an inclined electron crossing the 1 cm drift gap
at an angle of 0.1 rad (almost 6$^{\circ}$) with respect to the GEM foils.

The arrival time of the main clusters is clearly visible,
allowing an independent reconstruction of their 
absolute position in {\it z}.
Taking into account the gap width (1 cm)
and the width of the signal (about 135 ns),
an electron drift velocity of 72 $\mu$m/ns
is found in agreement with the value 
evaluated with Garfield.

\begin{figure}
\centering
\includegraphics[width=.55\textwidth, angle=90]{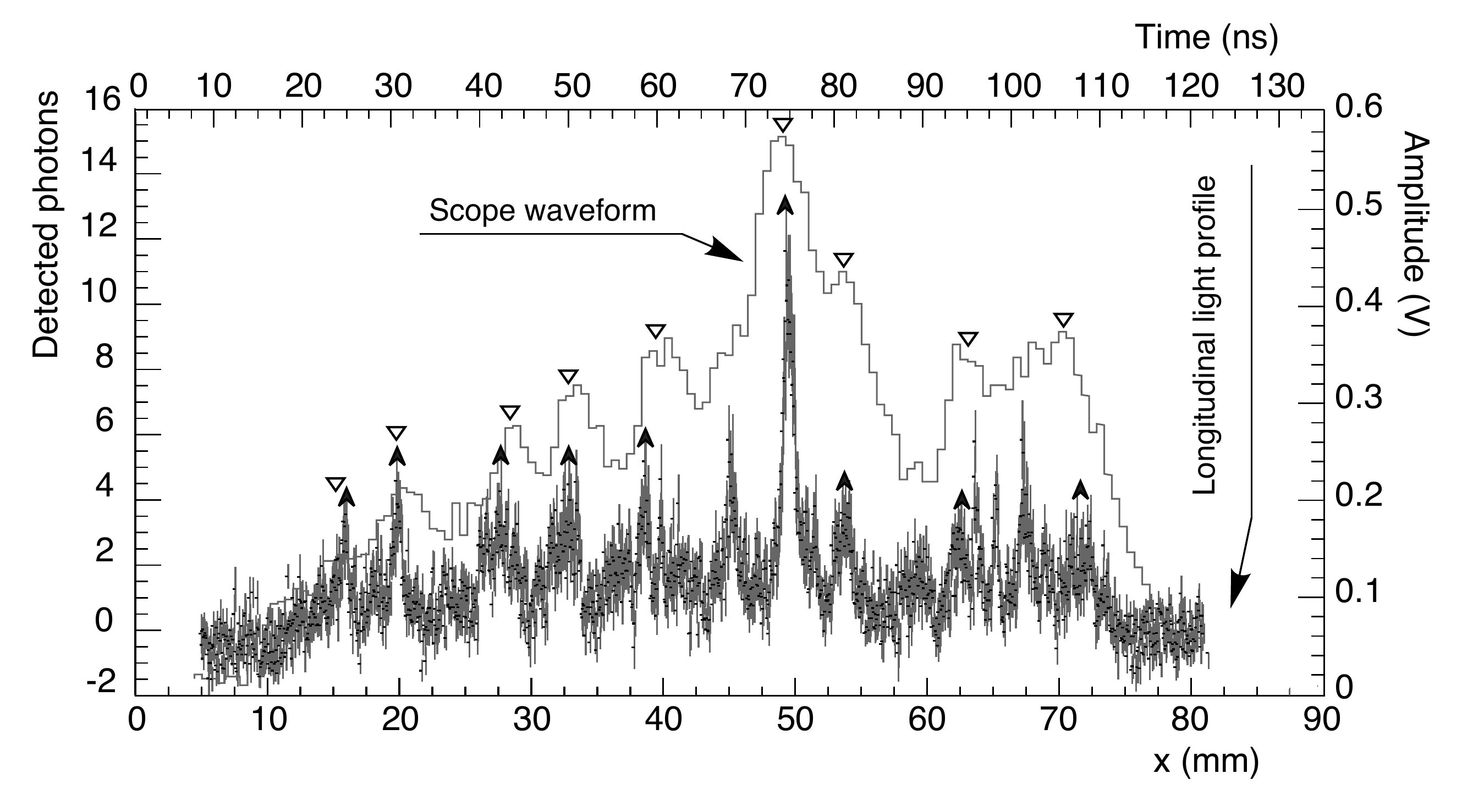}
\caption{Lateral profile of the light detected by the CMOS sensor along with 
the waveform of the PMT signal for the same event.
The cluster structure is clearly visible in both cases.
Peaks found by the finding algorithm are shown.}
\label{fig:peak}
\end{figure}

Figure \ref{fig:peak} shows an example of 
the lateral profile of the detected light 
as seen by the CMOS camera for a track
together with the corresponding PMT waveform.

In both cases, with a simple peak finding algorithm, 
the position of the main peaks was evaluated. 
The tracks in the ten analysed events 
have an average length of almost 60 mm
and 54 peaks were found in total. Therefore, the algorithm is
able to individuate one peak per track centimetre on average.

\begin{figure}
\centering
\includegraphics[width=.95\textwidth]{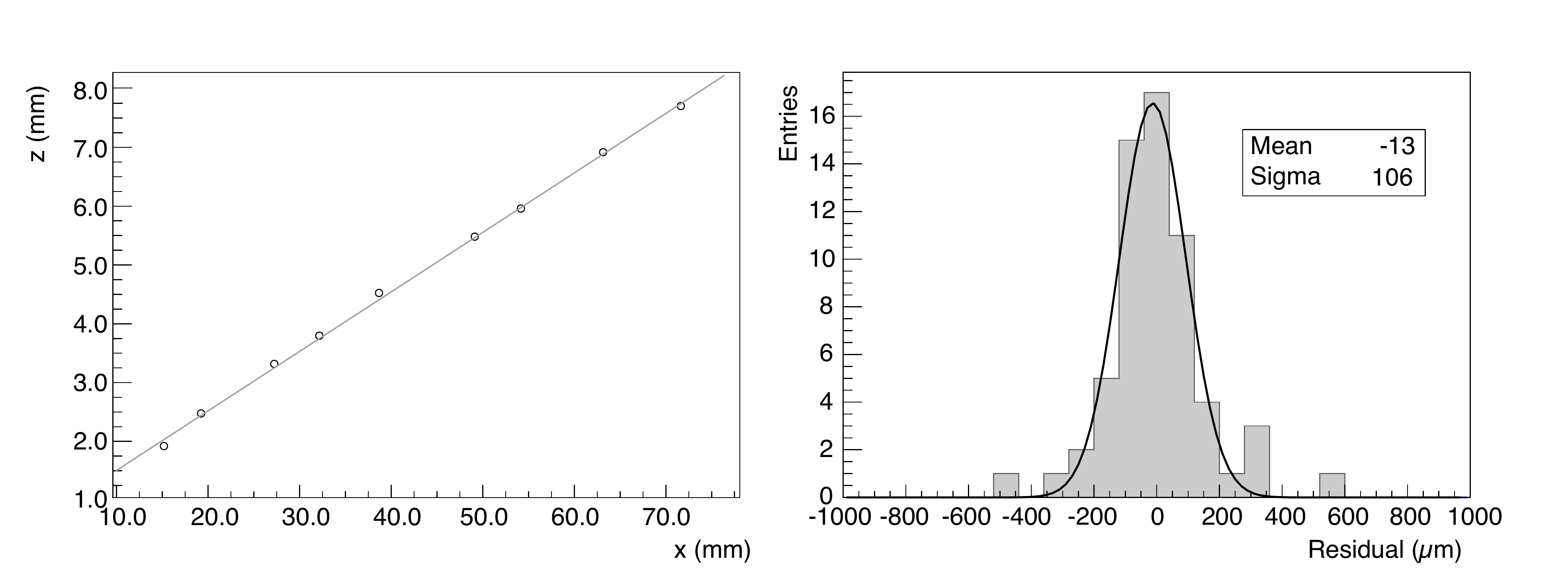}
\caption{
Left: Correlation with a superimposed linear fit of the
{\it x} and {\it z} coordinates of the clusters found in a single track.
Right: Distribution of the residuals of the reconstructed to the linear fit
{\it z} for ten tracks with a superimposed gaussian fit.}
\label{fig:zres}
\end{figure}

By assuming these peaks as due to ionization 
clusters along the tracks,
their {\it x} and {\it z} coordinates can be evaluated.
Their correlation is shown on the left of Fig. \ref{fig:zres}
while the distribution of the reconstructed
{\it z} residuals to a linear fit for a set of ten tracks
is shown on the right.

From the superimposed fit, it is possibile to evaluate a resolution 
on the reconstructed {\it z} coordinate of about 100 $\mu$m.

\section{Measurement of the released energy}

The light collected by the PMT is expected to be 
proportional to the energy released by the 450 MeV
electrons of the beam in the gas (i.e. 2.3 keV/cm
in an He/CF$_4$ 60/40 mixture \cite{bib:orange3}).
Before reaching the sensitive volume, electrons from the beam
lose energy in the material present on the beam line
as the black box containing the GEM structure or the
frame used to have a gas-tight volume.

\begin{figure}
\centering
\includegraphics[width=.95\textwidth]{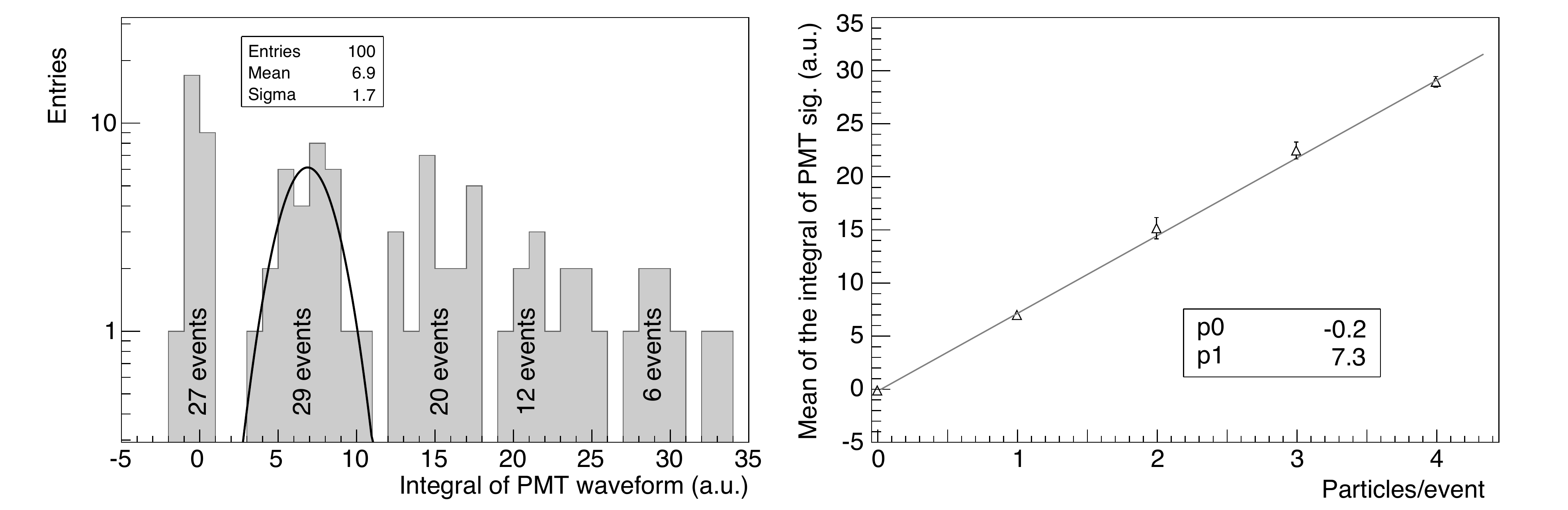}
\caption{
Left: Distribution of the integral of the PMT waveforms 
in a run of 100 events. For the
single-track events a superimposed gaussian fit is also shown.
Right: average PMT response as a function of the number of reconstructed particles
with a superimposed linear fit.}
\label{fig:nparticles}
\end{figure}

The distribution of the integral of the PMT signals
for 100 events
events is shown on the left of Fig. \ref{fig:nparticles}.
Separate peaks due to different numbers of
particles per event are clearly visibile.
The distribution of the number of
particles per event 
is compatible with a Poisson
distribution with an expected value of 1.4.

For the events with a single particle,
the jitter in the light detected by the PMT 
was studied as a reliable indicator of the resolution achievable in
the measurement of the released energy.

From the gaussian fit on the distribution for single particle event,
it is possible to evaluate that,
by reading out the light produced by means of a PMT, an energy
resolution of about 26\% can be easily achieved for an 
energy release of the order of 20 keV.

The average PMT response is shown on the right of Fig. \ref{fig:nparticles}
as a function of the number of particles. A very good linearity was found.
These last plots demonstrate that a simple light readout with a PMT allows
a fast evaluation of the number of particles crossing the sensitive volume
and the total energy released
that can be exploited, for example, for triggering purpose.

\section{Conclusion}

With the aim of constructing a high resolution GEM-based Time
Projection Chamber,
several R\&D activities are being carried on. 
In particular, the performance of a combined
optical readout (CMOS sensor and PMT) was studied in detail.
The photomultiplier signals showed a time resolution
better than 2 ns and demonstrated of being 
able to provide detailed information
on the time structure of the events, very useful for
a 3D reconstruction. A resolution of
100 $\mu$m on the coordinate orthogonal to the GEM plane was measured.
Moreover the analysis of the PMT waveform 
allows a fast measurement of the total released energy 
with an accuracy of 25\%.

\end{document}